\documentclass[aps,prb,twocolumn,superscriptaddress,10pt]{revtex4-2}
\usepackage[final]{graphicx}
\usepackage{sidecap}
\usepackage{wrapfig}
\usepackage[usenames,dvipsnames]{xcolor}
\usepackage{multirow}
\usepackage{mathtools}
\usepackage{fancyhdr}
\usepackage{braket}
\usepackage{cprotect}
\usepackage{tabularx}
\usepackage{empheq}
\usepackage{amsmath}
\usepackage{color, colortbl}
\usepackage[para]{footmisc}
\usepackage{footnote}
\usepackage[normalem]{ulem}
\usepackage{natbib}
\usepackage{stackengine}
\newcolumntype{M}[1]{>{\centering\arraybackslash}m{#1}}

\newcommand{\RomanNumeralCaps}[1]
    {\MakeUppercase{\romannumeral #1}}
    
\usepackage{hyperref}
\hypersetup{colorlinks=true, linkcolor=blue, citecolor=blue, urlcolor=blue}

\begin{document}
\widetext
\title{Origin of nonlinear magnetoelectric response in rare-earth orthoferrite perovskite oxides}

\author{Alireza Sasani}
\affiliation{Physique Th\'eorique des Mat\'eriaux, QMAT, CESAM, Universit\'e de Li\`ege, B-4000 Sart-Tilman, Belgium}

\author{Jorge I\~niguez}
\affiliation{MaterialsResearch and Technology Department, Luxembourg Institute of Science and Technology (LIST), 5 avenue des Hauts-Fourneaux, L-4362, Esch/Alzette, Luxemburg}
\affiliation{Department of Physics and Materials Science, University of Luxembourg, Rue du Brill 41, L-4422 Belvaux, Luxembourg}

\author{Eric Bousquet}
\affiliation{Physique Th\'eorique des Mat\'eriaux, QMAT, CESAM, Universit\'e de Li\`ege, B-4000 Sart-Tilman, Belgium}

\begin{abstract}
We report a theoretical study of the non-linear magnetoelectric response of GdFeO$_3$ through an analytical approach combined with a Heisenberg model which is fitted against first-principles calculations. Our theory reproduces the non-linear change of polarization under applied magnetic field reported experimentally such that it allows to analyze the origin of the large responses in the different directions.  
We show that the non-linear character of the response in these materials originates from the fact that the antiferromagnetic order of Gd atoms changes non-linearly with respect to the applied magnetic field. Our model can be generalized to other materials in which the antiferromagnetic ordering breaks inversion symmetry. 
\end{abstract}

\maketitle

\newpage

\section{Introduction}
Magnetoelectric (ME) materials are compounds in which there is a coupling between magnetic field (magnetization) and electric field (polarization)~\cite{Landau-1982, curie_1894}.
Magnetoelectricity is a sough after materials response because it allows to control the magnetic properties using an external electric field or, the other way around, to control the polarization using an external magnetic field and, hence, it can have plethora of possible applications in spintronics, sensors, etc~\cite{Brown-1968, Dong2015, Spaldin2019}. In particular these materials can be used for memory devices by enhancing the speed of the device performance together with reducing its energy consumption \cite{Spaldin2019,Spaldin2020}.

Since the first experimental observation of the ME effect by Astrov \cite{Astrov1960} there have been many works to find the ME effects in other materials but it appeared that most of these ME responses were very small to be considered practical~\cite{Fiebig2005,Dong2015, Spaldin2019, Wang2009}. 
So far the discovered ME materials have either a small coupling \cite{Wang2003}, or a very low performance temperature  \cite{Kimura2003} which hinders putting them into applications. So far different paths were introduced to enhance the ME response of  materials. To name a few, we have structural softness through epitaxial strain~\cite{Wojdet_2010,Bousquet2011} or solid solutions~\cite{Bayaraa2021} or making artificial structures and superlattices \cite{Zhao2014, Cherifi2014}.

Bulk multiferroic (MF) materials are a subclass of MEs in which there exists two ferroic orders in the same phase, i.e., ferroelectricity and ferromagnetism (or antiferromagnetism)\cite{Schmid_1994}. 
This class of materials are divided into two groups, namely, type-\RomanNumeralCaps {1} and type-\RomanNumeralCaps {2}~\cite{Khomskii2009}. Type-\RomanNumeralCaps {2} MFs are materials in which the magnetic ordering is the mechanism that breaks the inversion symmetry causing ferroelectricity.
Hence, in type-\RomanNumeralCaps {2} MFs a strong coupling between magnetism and polarization is present by construction, resulting in strong ME responses with respect to type-\RomanNumeralCaps {1} MFs where the coupling is more indirect.
The reported ME responses for these materials show that indeed the strongest ME responses are found in type-\RomanNumeralCaps {2} MFs~\cite{Chapon2004,Kenzelmann2005, Kimura2003, Tokunaga-2009, Tokunaga-2008}.
In a type-\RomanNumeralCaps {2} MF, the ME response can result from either non-collinear spin ordering, in which we expect small polarization due to the fact that it arises from spin orbit coupling (SOC) (10$^{-2} \mu C/cm^2$ ); or it can result from inversion symmetry breaking due to collinear ordering of the spins. In the latter case, the mechanism can be other than SOC,  like e.g. exchange striction, which typically yields large polarization (one to two orders of magnitude larger) compared to the first mechanism.
Some of the rare earth orthoferrites (e.g. GdFeO$_3$) are of type-\RomanNumeralCaps {2} multiferroics, in which the collinear ordering of spins creates the polarization. Hence, they have larger polarization compared to other type-\RomanNumeralCaps {2} MFs as well as larger ME responses.
Although the temperature at which the multiferroicity appears is very low (it requires that the rare earth spins become ordered), their ME responses are among the highest reported in single phase materials\cite{Tokunaga-2009,Tokunaga-2008} and, regarding the amplitude, they can compete with laminated composites (sandwiched structures of piezoelectric and magnetostrictive materials in which the strain coupling between the two materials is the mechanism creating large ME response)\cite{Ryu2002,Vaz_2010}.
These rare earth orthoferrite materials have such a strong coupling that makes it possible to control ferroelectric order using magnetic fields or to control the magnetic ordering using electric fields \cite{Tokunaga-2009}.
However, the exact origin of their large responses has not been fully analyzed from first-principles or model Hamiltonian (fitted to first-principles) calculations due to the complexity associated to the presence of two different and coupled magnetic sublattices, and because the rare earth magnetism comes from $f$ electrons which are difficult to handle in density functional theory (DFT) calculations \cite{Stroppa2010,Chen2012}. 

In this paper we report a simulation study of the microscopic origin of the large non-linear ME response of GdFeO$_3$. 
GdFeO$_3$ is among the largest reported ME response in single crystals. Indeed, experimentally it has been observed that the polarization can go from 0.14 $\mu C/cm^2$ to 0 $\mu C/cm^2$ under an applied magnetic field of 4 T~\cite{Tokunaga-2009}. 
If we extrapolate an effective linear response in the unit of ps/m by calculating $\frac{\Delta P}{\Delta B}$ between 0 and 4 T, we obtain an effective amplitude of about 4600 ps/m, i.e about seven times larger than the linear ME crystal TbPO$_4$ (~730 ps/m largest linear ME response)\cite{Rivera_2009}.
To tackle this problem from a simulation view point, we first derive an analytical form of the ME response of this material using both a Heisenberg Hamiltonian and DFT calculations to fit the parameters. 
Then, we report the results obtained through a classical spin dynamics to calculate the ME response and the polarization of these materials under an applied magnetic field. 
Our results reproduce the response observed experimentally on GdFeO$_3$, i.e., the fully non-linear response and the appearance of two regimes, associated to a magnetic phase transition under the applied magnetic field.

\section{Technical details}
In this work we have used the Heisenberg model that has been derived previously by us in Ref.~\cite{sasani2021magnetic}, which includes both rare earth and transition metal site interactions (superexchange and Dzyaloshinskii-Moriya interactions).
This model is fitted against density functional theory (DFT) calculations~\cite{Hohenberg1964,Kohn1965} of the $Pna2_1$ phase of GdFeO$_3$.
We used the VASP DFT package~\cite{KRESSE199615,Kresse-1996} and its projected augmented wave implementation~\cite{Blochl1994}.
We used the so-called generalized gradient approximation (GGA) of the PBEsol flavor~\cite{Perdew-2008} for the exchange correlation functional and added a Hubbard $U$ correction~\cite{Liechtenstein1995} on Fe and Gd of respectively 4 eV and 5 eV as well as a $J$ parameter of 1 eV on Fe. Since the behavior that we are interested in is arising from exchange interactions, we have chosen Hubbard interaction parameters ($U$) so that we get the closest N\'eel temperature compared to experiments. All the calculations were done considering non-collinear magnetism and including the spin orbit interaction. 
The calculations were converged with a 6$\times$6$\times$4 mesh of k-points for sampling the reciprocal space and a cut-off energy on the plane wave expansion of 700 eV (giving a precision of less than 5 $\mu\text{eV}$ on the single-ion anisotropy and the DMIs).

The calculations of the superexchange interactions were done using Green's function method as implemented in the TB2J~\cite{he2020tb2j} code. In this method the maximally localized Wannier function~\cite{Marzari-1997} as implemented in WANNIER90~\cite{MOSTOFI20142309} are calculated using DFT (VASP interface to Maximally localized Wannier functions) and, using these Wannier functions and the Green's function method, the superexchange parameters are calculated.
Some of these superexchange interactions were compared to the ones calculated using total energy to ensure the consistency of the method. 
To determine the DMI amplitudes, we have calculated the energy of different spin configurations as described by Xiang {\it et al}.~\cite{xiang_magnetic_2013}.
All of the fitted magnetic interaction parameters were used to do spin dynamics with the VAMPIRE code~\cite{evans2014}. In this code the Landau-Lifshitz-Gilbert (LLG) equation for the spin dynamics
\begin{equation} \label{eq:llg}
\begin{array}{l@{}l}
\frac{\partial S_i}{\partial t}=\frac{\gamma}{1+\lambda^2}\left[S_i\times B_{eff}^i + \lambda S_i \times \left(S_i \times B_{eff}^i \right)  \right]\\
\end{array}
\end{equation}
is solved numerically. The ground state (lowest energy solution at T=0 K) for each of the calculations (with applied magnetic field) is found by minimizing the energy with respect to the magnetic order .

The calculation of the polarization is done using the Berry phase approach as implemented in VASP \cite{Smith_1993,Resta_1994}. 
In order to calculate the ferroelectric response as a function of magnetic order in the structure, we have constrained magnetic order and rotated the spins on rare earth site from antiferroamgnetic order to feromagnetic order and relaxed the atomic structure at each step. To make a model to simulate the ME response of the materials we have fitted the spin dynamics results with the models that we have developed in the following.

Before analyzing the results obtained with the fitted model, we start with an analytical discussion of the magnetic interactions present in GdFeO$_3$.


\section{theoretical derivation}
In this section, we derive an analytical expression of the ME response of the GdFeO$_3$ originating from the exchange striction interaction.
To derive the ME response, we use the fact that the $G$-type (or $A$-type) magnetic order of the rare earth site is the driving force that breaks the inversion symmetry and causes the polarization in rare earth orthoferrites~\cite{Tokunaga-2009,Bertaut,Zhao2017}.
Hence, we will consider the polarization to be a function of the rare earth site $G$-type order such that we can expand the polarization in terms of the corresponding order parameter. From this assumption, we can write the ME response in these structures using the following relations (Einstein summation rule applies):
\begin{equation} \label{eq:1}
\begin{array}{l@{}l}
\frac{\partial P_i}{\partial B^{app}_l}=\frac{\partial P_i}{\partial G^{R}_j}\frac{\partial  G^{R}_j}{\partial B^{app}_l},
\end{array}
\end{equation}
where $ B^{app}_{l}$ is the applied magnetic field in the $l$ direction, $P_i$ is the polarization in the $i$ direction and $G^{R}_j$ is the magnitude of the G-type order on rare earth site in the $j$ direction.

To probe the variation of G-type order with respect to the magnetic field, $\frac{\partial  G^{R}_j}{\partial B^{app}_l}$, we use the general Heisenberg model developed by us for $RM$O$_3$ crystals ($R=$ Rare-earth, $M=$ Fe or Cr) \cite{sasani2021magnetic} but using the data fitted from DFT calculations done on GdFeO$_3$. 
According to our simulations the magnetic order on the Fe sublattice is not affected by the range of magnetic fields that we are interested in, so we will neglect the Fe sublattice interaction in our theoretical derivations (see Fig. \ref{fig:Fe_z}). 
In this approximation, the energy of rare earth spins per formula unit can be derived from the Heisenberg Hamiltonian to obtain:
\begin{widetext}
\begin{equation} \label{eq:2}
\begin{array}{l@{}l}
H_{Heis}^{R}=-3 J^R (G^{R}_i)^2 + 3 J (F^{R}_j)^2 - K^R (G^{R}_i)^2 - B^{app}_l F^{R}_j \delta_{jl} - B^{RM}_m F^{R}_j \delta_{mj},
\end{array}
\end{equation}
\end{widetext}
where $J^R$ is the exchange interaction between rare earth sites and $G^{R}_i$ and $F^{R}_j$ are the G-type antiferromagnetic (AFM) and ferromagnetic (FM) orders on rare earth in the $i$ and $j$ directions respectively, $K$ is the single ion anisotropy in the $i$ direction. 
The interaction of Gd and Fe spins can be reduced to an effective magnetic field $B^{RM}_m$, which can be written as follows~\cite{Zhao-2016,sasani2021magnetic}:
\begin{equation} \label{eq:3}
\begin{array}{l@{}l}
B^{RM}_m=8J^{RM} F^M_m + 8(d^{RM} \times G^M)_m,
\end{array}
\end{equation}
 where $J^{RM}$ is the exchange interaction between rare earth and transition metal and the $d^{RM}$ is the Dzyaloshinskii-Moriya interaction (DMI) between rare earth and transition metal cations. In $d^{RM} \times G^M$ term only the y component of the DMI (which is also the largest component~\cite{sasani2021magnetic}) produces an effective magnetic field and in the cross product we only consider the  term corresponding to this component. We consider the effective magnetic field $B^{RM}_m$ to be in the same direction as applied magnetic field, or small compared to it such that it can be neglected. In the case where the applied field is in the $z$ direction and the rare earth orders in the G type in $x$ direction, the $B^{RM}_m$ and the applied magnetic field are in the same direction~\cite{sasani2021magnetic, Zhao-2016}. 
 When the applied field is in the $x$ direction, $B^{RM}_m$ and the applied field are perpendicular to each other before the phase transition (in which we can consider $B^{RM}_m$ to be negligible compared to the applied magnetic field), while after the phase transition they will be in the same direction.\\
 In this case in our Heisenberg Hamiltonian the total magnetic field acting on the rare earth site can be written as:
\begin{equation} \label{eq:2-1}
\begin{array}{l@{}l}
B_{l}^{eff}F^{R}_l=( B^{app}_l + B^{RM}_l )F^{R}_j  \delta_{lj}
\end{array}
\end{equation}

We can minimize the energy with the following constraint using Lagrange multipliers (the constraint is coming from considering the magnitude of the spin as normalized to one):
\begin{equation} \label{eq:4}
\begin{array}{l@{}l}
(G^{R}_i)^2+(F^{R}_j)^2=1,
\end{array}
\end{equation}
which gives us the $G_i$ and $F_j$ orders as a function of the applied magnetic field:
\begin{equation} \label{eq:5}
\begin{array}{l@{}l}
G^{R}_i=\pm\sqrt{1-  \left( \frac{B_{l}^{eff}}{12J^R+2K^R} \right)^2}
\end{array}
\end{equation}
\begin{equation} \label{eq:6}
\begin{array}{l@{}l}
F^{R}_l=\frac{B_{l}^{eff}}{12J^R+2K^R}.
\end{array}
\end{equation}
From these expressions we can obtain the following term:
\begin{equation} \label{eq:7}
\begin{array}{l@{}l}
\frac{\partial  G^{R}_j}{\partial B^{app}_l} =\mp\frac{\frac{B_{l}^{eff}}{(12J^R+2K^R)^2}}{\sqrt{1-\left( \frac{B_{l}^{eff}}{12J^R+2K^R} \right)^2}}.
\end{array}
\end{equation}
This equation holds as long as there is no magnetic phase transition in the structure.

Now, we are left with the determination of the variation of $P$ with respect to $G^{R}_j$: $\frac{\partial P_i}{\partial G^{R}_j}$.
Because the exchange striction between the rare earth site and the transition metal site is the interaction responsible for polarizing the material~\cite{Tokunaga-2009,Zhao2017} we are going to use the energy expression for this interaction to derive the $\frac{\partial P_i}{\partial G^{R}_j}$. We can write the energy from exchange interaction between rare earth site and transition metal site as:

\begin{equation} \label{eq:8}
\begin{array}{l@{}l}
E_{int}=4 G^{R}_i G^{M}_j(J^{RM^+}_{ij}-J^{RM^-}_{ij})
\end{array}
\end{equation}
where we have separated the exchange interaction for up and down directions of the transition metals spins ($ J^{RM^+}_{ij}, J^{RM^-}_{ij}$) with respect to the rare earth spins. 
We can notice that this interaction is zero in centrosymmetric structures since the exchange for up and down directions of the spins are the same. 

If we define $J_{ij}^{rm}=J^{RM^+}_{ij}-J^{RM^-}_{ij}$ as a change in exchange interaction which results in breaking of inversion symmetry, we can derive the polarization as:

\begin{equation} \label{eq:8.5}
\begin{array}{l@{}l}
P_k=\frac{\partial E_{int}}{\partial \vec{\varepsilon}_k}=\frac{\partial E_{int}}{\partial J_{ij}^{rm}}.\frac{\partial J_{ij}^{rm}}{\partial \vec{\varepsilon}_k},
\end{array}
\end{equation}
where $\vec{\varepsilon}_k$ is the electric field in $k$ direction.
The first part of the derivation can be derived from eq.\ref{eq:8} as:
\begin{equation} \label{eq:9}
\begin{array}{l@{}l}
\frac{\partial E_{int}}{\partial J^{rm}_{ij}}=4 G^{R}_i G^{M}_j,
\end{array}
\end{equation}

Since the change in exchange interaction in very small, we are going to consider an expansion in terms of electric field up to linear order. So, we can have $\frac{\partial J_{ij}^{rm}}{\partial \vec{\varepsilon}_k}$ as: 
\begin{equation} \label{eq:10}
\begin{array}{l@{}l}
 \frac{\partial J_{ij}^{rm}}{\partial \vec{\varepsilon}_{k}} = \delta_k^{ij}
\end{array}
\end{equation}
Where $\delta_k^{ij}$ presents change in exchange interaction $J_ij$ due to displacement of the atoms in $k$ direction.
From now on, we consider the symmetric exchange interactions ($j=i$ in eq. \ref{eq:8}) between $R$ and $M$, since the antisymetric interactions (DMI) are already included as $B^{RM}_{j}$ (we have neglected the changes in DMI interactions due to lattice distortions).
We neglect also the anisotropic symmetric exchange since these interactions are negligible compared to the other interactions in GdFeO$_3$~\cite{sasani2021magnetic}.
Hence, for symmetric exchange interaction we can write:
\begin{equation} \label{eq:11}
\begin{array}{l@{}l}
P_i=4\delta_i^{jj} G^{R}_j G^{M}_j 
\end{array}
\end{equation}

  \begin{figure}[htb!]
  	\centering
  	\includegraphics[width=1\linewidth ,keepaspectratio=true]{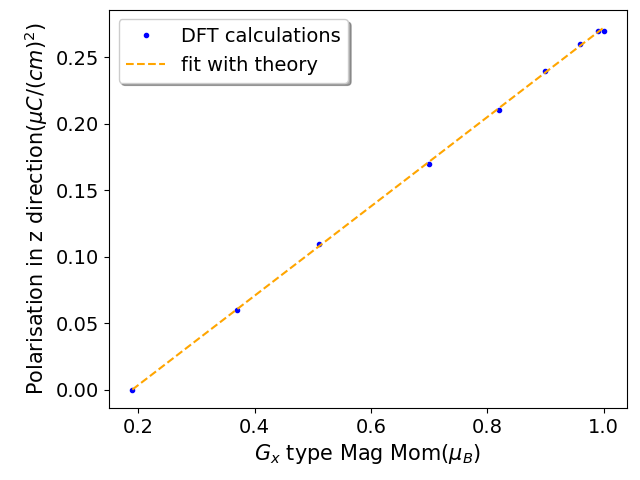} 
  	\caption{Polarization of the GdFeO$_3$ as a function of magnitude of G-type order in the material. We also have shown the fitted line with the theory}
  	\label{fig:pol_dft}
  \end{figure}
\noindent  
In this case we have:
\begin{equation} \label{eq:12}
\begin{array}{l@{}l}
\frac{\partial P_i}{\partial G^{R}_j}=4 \delta_i^{jj} G^M_j,
\end{array}
\end{equation}

To determine the strength of the change of polarization as a function of the magnitude of the G-type AFM ordering of rare earth ( as Eq. \ref{eq:12}), we performed DFT calculations. 
We have calculated the polarization for different magnetic ordering by changing the spin order from G-type order to ferromagnetic order by rotating it slowly.
In Fig.~\ref{fig:pol_dft} we report the results, i.e., the change of polarization as a function of the G-type order magnitude as we go from G-type to the ferromagnetic order. 
We can notice the linear relation between G-type order magnitude and the polarization, which proves that the Eq.~\ref{eq:11} is a good estimation of the polarization of the materials as a function of G-type order magnitude.

By fitting the Eq.~\ref{eq:12} with the results of Fig.~\ref{fig:pol_dft}, we can extract the coefficients of this equation.
We obtain a slope of 0.336 $\mu C cm^{-2}\mu_B^{-1} $ with the polarization in the $z$ direction which is perpendicular to $G_x$ type magnetic ordering (from now on we will consider the polarization in the $z$ direction). Hence we have:
\begin{equation} \label{eq:18}
\begin{array}{l@{}l}
\frac{\partial P_z}{\partial G^{R}_x}=4 \delta_z^{xx} G^M_x= 0.336~\mu C cm^{-2}\mu_B^{-1}
\end{array}
\end{equation}
for GdFeO$_3$. 
With this coupling term at hand, we can explore how the crystal responds to an applied magnetic field with the Heisenberg model and deduce how the polarization changes, i.e. the ME response. We use the same applied field conditions as reported experimentally for GdFeO$_3$ by Tokunaga \textit{et. al.} ~\cite{Tokunaga-2009}.

We can now have following analytical expression for the ME response:
\begin{equation} \label{eq:15}
\begin{array}{l@{}l}
\frac{\partial P_z}{\partial B_{l}^{app}}= \mp\frac{4 \delta_z^{jj} G^M_j \frac{B^{eff}_l}{(J_P)^2}}{\sqrt{1-\left( \frac{B^{eff}_l}{J_P} \right)^2}}
\end{array}
\end{equation}

If we consider $B^{eff}_l > J_p$ with $J_p=12J^R+2K^R $ and where the system is not completely ferromagnetically ordered, which is a good assumption since we are studying the response of the system in this regime, we can expand this function around zero applied magnetic field as follows:
\begin{widetext}
\begin{equation} \label{eq:17}
\begin{array}{l@{}l}
\frac{\partial P_z}{\partial B_{l}^{app}}=\mp\frac{4 \delta_z^{jj}  G^M_j \frac{B^{eff}_l}{(J_p)^2}}{\sqrt{1-\left( \frac{B^{eff}_l}{J_p} \right)^2}} = \mp 4 \delta_z^{jj} G^M_j \frac{B^{eff}_l}{(J_p)^2}(1+\frac{1}{2}\left( \frac{B^{eff}_l}{J_p} \right)^2+\frac{3}{8}\left( \frac{B^{eff}_l}{J_p} \right)^4+...),
\end{array}
\end{equation}
\end{widetext}
where the negative and positive signs are for the positive and negative direction of the applied magnetic field respectively.

We can see that the ME response is non-linear because the AFM order changes non-linearly under an applied magnetic field and we can expect this non-linear behavior for all the cases where the AFM order breaks the inversion symmetry (this should be the case for A type AFM order and E-type AFM order).
While the AFM order creates a non-linear ME response, the FM order that drives ferroelectricity will have a linear ME response before magnetization saturation (since the ferromagnetic order changes linearly with respect to applied magnetic field, see Eq. \ref{eq:6}). 
Another point to mention is the fact that the A-type AFM ordering of the rare earth site can also break the inversion center in these structures and can induce nonlinear polarization.
If we consider the ME response from this ordering we should change the denominator $12 J^R $ by $ 4 J^R $ in the ME response.

If we integrate Eq.~\ref{eq:15} with respect to the magnetic field, we can calculate the polarization as a function of the magnetic field for these materials using the initial values obtained from DFT.
This integration gives the following final analytical expression:
\begin{widetext}
\begin{equation} \label{eq:Pola}
\begin{array}{l@{}l}
P_z(B^{app}_l)= P_z(B^{app}_l=0)- 4 \delta_z^{jj} G^M_j \left(\sqrt{1- \left( \frac{B^{eff}_l}{J_P} \right)^2} \right)
\end{array}
\end{equation}
\end{widetext}

Now that we have analyzed analytically the ME response of rare earth perovskites in the magnetic phases as present in ferrites and chromites, in the next section we present our numerical results coming from the simulations for GdFeO$_3$.


\section{simulations}
In this section we present the results of the spin dynamics simulations to study numerically the effect of the applied magnetic field on magnetism and the resulting ME response.

\subsection{Magnetic fields perpendicular to the Gd G-type order direction}

In this part we discuss the ME response of GdFeO$_3$ as a function of magnetic field applied perpendicular to the direction of the Gd spins with G-type order.

\begin{figure}[htb!]
  	\centering
  	\includegraphics[width=1\linewidth ,keepaspectratio=true]{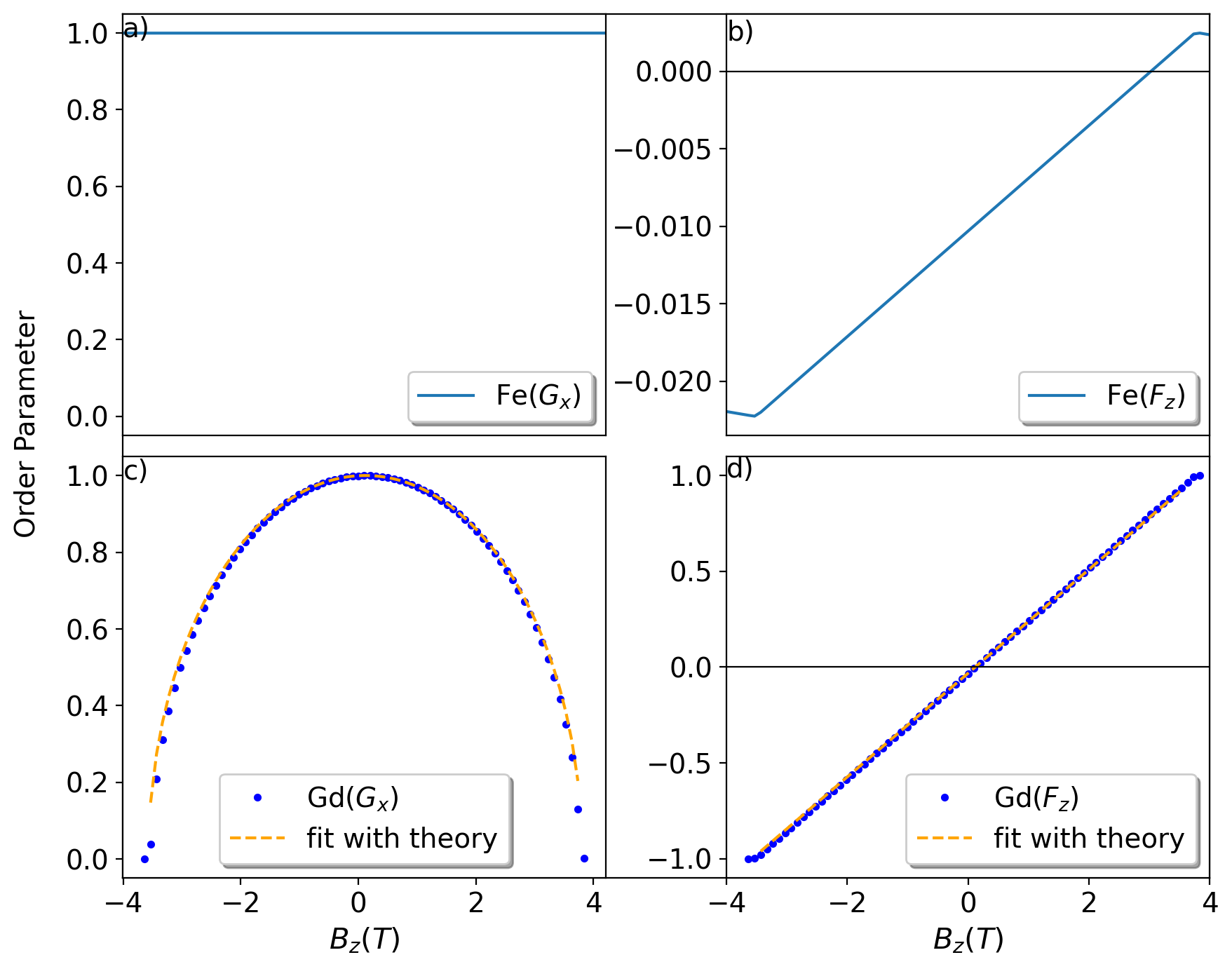} 
  	\caption{Fe and Gd site magnetic ordering for spins in x and z direction as a function of applied magnetic field in z direction.}
  	\label{fig:Fe_z}
  \end{figure}

In Fig.~\ref{fig:Fe_z}-a and Fig.~\ref{fig:Fe_z}-b we show the spin dynamics results of the effect of an external magnetic field on Fe sublattice. We notice that the Fe sublattice does not change much as the magnetic field is applied to the structure, we can only observe a small change in its weak ferromagnetic canting.
This result shows that we can neglect the Fe magnetic order changes effects on the ME response since the effects for Gd sublattice are much larger.

In Fig.~\ref{fig:Fe_z}-c and Fig.~\ref{fig:Fe_z}-d we report the effect of the applied magnetic field in the $z$ direction on the Gd sublattice. We can see a decrease of the $G$ type ordering along the $x$ direction and an increase of the $F$ type along the $z$ direction. Hence, the applied magnetic field can fully magnetize the Gd parallel to the field direction.
Beyond a critical field of 4 T, we can see that the ground state $G$ type AFM order has disappeared, the magnetic field having fully magnetized all the Gd moment in the same direction.
This transition is fully consistent with the experimental results of Ref.~\cite{Tokunaga-2009}. 

To check the consistency of the spin dynamics results with the analytical solution that we have obtained in the previous section, we fitted the results of the ferromagnetic order ($z$ component of the Gd spin) with Eq.~\ref{eq:6} to obtain the $J^R$, $K^R$ and $B_l^{RM}$ parameters.
The orange dashed line in Fig.~\ref{fig:Fe_z}-d  shows the resulting fit that is in good agreement with the spin dynamics results ( blue dots). We then used these parameters and put them in Eq.~\ref{eq:5} and plotted the results in Fig.~\ref{fig:Fe_z}-c for the $x$ component of the Gd spin.
The values for the parameters obtain from the fit with spin dynamics are close to the values calculated from DFT. As we can see these results agree well with the spin dynamics simulations.

\begin{figure}[htb!]
  	\centering
  	\includegraphics[width=1\linewidth ,keepaspectratio=true]{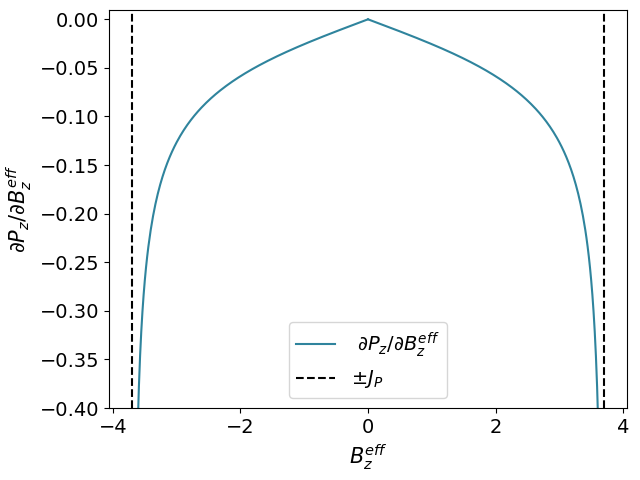} 
  	\caption{Non-linear Magneto electric response of GdFeO$_3$ orthofrrites plotted according to Eq. \ref{eq:15} where we can see a divergence in the response as applied magnetic field strength get closer to $J_p$ }
  	\label{fig:ME_RESP}
  \end{figure}

Having calculated the required coefficients for the magnetic response, we can calculate the ME response. 
In Figure~\ref{fig:ME_RESP} we report the evolution of the change of polarization versus the applied magnetic field.
We can see that the change of polarization response is negative (i.e. the magnetic field reduces the polarization), symmetric for each magnetic field directions and diverges when approaching the critical field where the Gd order goes from $G_x$ to $F_z$.
This critical field is directly related to the amplitude of the Gd exchange interaction ($J^R$), which governs the energy change associated to the change of the Gd magnetic order (here from $G$ to $F$ type).
This corresponds to the phase transition from the polar $m'm2'$ ($Pna2_1$) phase to the paraelectric $m'm'm$ ($Pnma$) phase 

\begin{figure}[htb!]
  	\centering
  	\includegraphics[width=1\linewidth ,keepaspectratio=true]{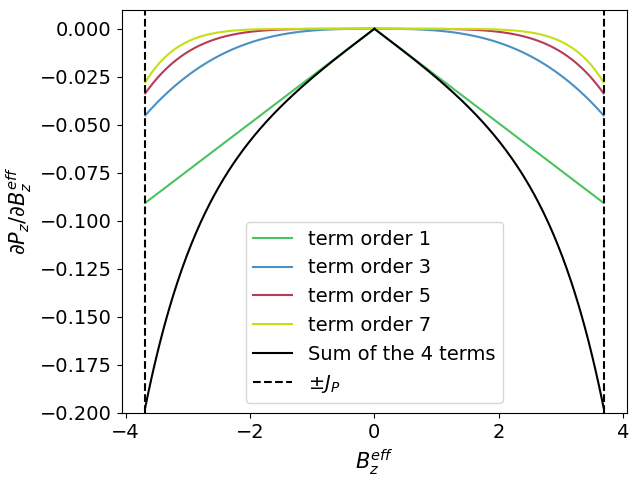} 
  	\caption{Non-linear Magneto electric response of Gd orthofrrites plotted according to Eq. \ref{eq:17} for different orders of the equation where we can see that the divergence is mainly coming from higher order terms in  response of the material} 
  	\label{fig:MERs_ordrs}
  \end{figure}

In Fig.~\ref{fig:MERs_ordrs} we report the ME response decomposed into its different expansion orders as made in Eq.~\ref{eq:17}. 
We can see that close to zero magnetic field the response is mainly driven by its linear term (second order ME response) but as we are going with higher magnetic field amplitudes the higher order non-linear responses become more and more important. 

  \begin{figure}[htb!]
  	\centering
  	\includegraphics[width=1\linewidth ,keepaspectratio=true]{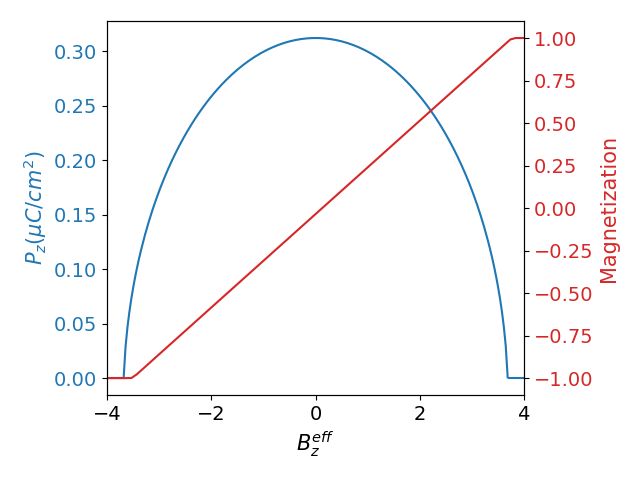} 
  	\caption{polarization as a function of the magnetic field for GFO where it reaches to zero as the applying magnetic field strength reaches that of exchange interaction between Gd ($J^R$), and magnetization of the Gd (red line)  }
  	\label{fig:Pola_VS_H}
  \end{figure}

In Fig.~\ref{fig:Pola_VS_H} we show the polarization versus the magnetic field of GdFeO$_3$ as obtained from Eq.~\ref{eq:Pola}. 
We obtain that the polarization decreases non-linearly for both direction of the field and reaches zero at the critical magnetic field where the crystal goes to the $Pnma$ paraelectric phase.
In this figure we have also included the magnetization of the crystal coming from the Gd where we can see that when the the Gd sublattice if ferromagnetically ordered, the polarization vanishes.
This result is in very good agreement with the experimental results of Tokunaga et al. \cite{Tokunaga-2009}, which also proves that our model describes correctly the ME response of this material.


 \subsection{ME response for magnetic field parallel to the Gd G-type order:}

\begin{figure}[htb!]
	\centering
	\includegraphics[width=1\linewidth ,keepaspectratio=true]{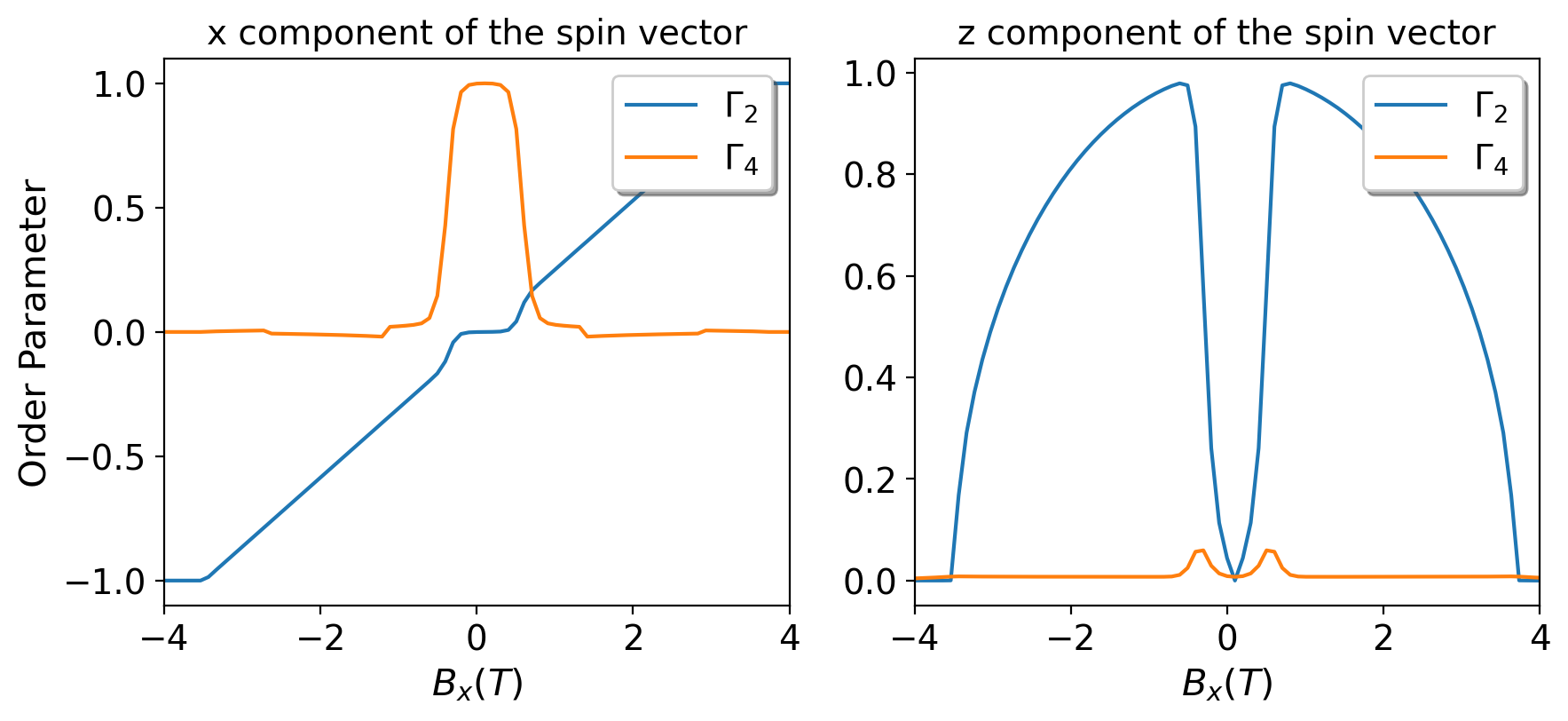} 
	\caption{Re site magnetic ordering as a function of applied magnetic field in x direction. For the magnetic field of less than 1T we can see the sudden drop of G-type order in x direction (the orange line in x component of spin) which is accompanied by a increase in G-type order in z direction (blue line in z component of spin)}
	\label{fig:Re_x}
\end{figure}

 In this section we discuss the ME response of GdFeO$_3$ for an applied magnetic field parallel to the the direction of the Gd $G$-type magnetic order.
Similarly to the applied field along the $z$ direction, our simulations give that the magnetic ordering of the Fe site is not strongly affected by the applied magnetic field along the $x$ direction such that we can neglect it.
In Fig.~\ref{fig:Re_x} we report the evolution of the Gd sublattice spin when we apply a magnetic field in $z$ direction.
We can see that for a critical magnetic fields of about 1 T, the Gd goes thorough a spin-flip phase transition from the $\Gamma_4$ state with $G$-type order in $x$ direction (orange line) to the $\Gamma_2$ state with the $G$-type order along the $z$ direction (blue line). 
In this structure the Gd atoms prefer to be in $\Gamma_4$ state due to single ion anisotropy and also effective field of Fe. By applying magnetic field in x direction we lower the energy of the $\Gamma_2$ state by $B_x.F_x$ (where $F_x$ is a weak canted moment characteristic of the $\Gamma_2$ state) and once this energy is larger than the energy difference between $\Gamma_4$ and $\Gamma_2$ we will have a phase transition.
Beyond this phase transition, the Gd spins start to be more and more ferromagnetic and it becomes fully magnetized for the amplitude of 4~T magnetic filed.

\begin{figure}[htb!]
  	\centering
  	\includegraphics[width=1\linewidth ,keepaspectratio=true]{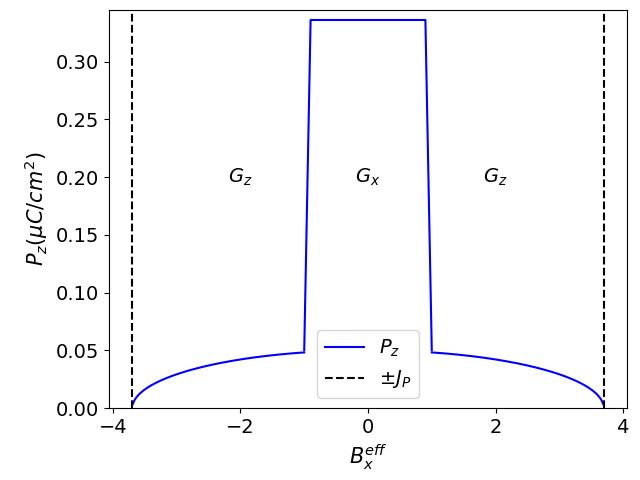} 
  	\caption{change of polarization for as a function of applied magnetic field in x direction where we can see the ordering of R site in $G_x$ before phase transition and in $G_z$ with lower polarization}
  	\label{fig:Pola_VS_Hx}
  \end{figure}

As done previously for the applied field along the $z$ direction, we can also calculate how the polarization if affected by the applied field along the $x$ direction and so the ME response.
In Fig.~\ref{fig:Pola_VS_Hx} we report the evolution of the polarization versus the applied magnetic field along $x$.
We encounter a non-linear ME response again where the polarization is decreased for both directions of the field.
We, however, observe two regimes, one between 0 and $\pm$1T where the polarization is approximately constant and not affected by the field. Exactly at $\pm$1T we observe a sharp polarization drop (from 0.26 $\mu C/cm^2$ to 0.05 $\mu C/cm^2$, a reduction by a factor of 5) due to the transition of Gd sublattice from the $\Gamma_4$ to the $\Gamma_2$ phase .

Then, beyond $\pm$1T we have a non-linear further reduction of the polarization down to zero when the Gd is fully magnetized by the field along the $x$ direction.
The polarization will change like Eq.~\ref{eq:Pola} for the range of fields between 1 T and 4 T with a different exchange coupling: in the first case (where we applying the magnetic field in z direction) 
we had both Gd and Fe atoms spin in x direction and we were considering $\delta_{z}^{xx}= J^{RM^+}_{xx}-J^{RM^-}_{xx}$; instead, now we will have $\delta_z^{zx}= J^{RM^+}_{zx}-J^{RM^-}_{zx}$ (Fe in G-type ordering in x direction and Gd in G-type ordering in z direction) which is much smaller than the the first $\delta_z$ hence resulting in a smaller polarization for this part.
The ME response will also be similar to Eq.~\ref{eq:15} for magnetic fields higher than 1 T.

Again, our results reproduce well the experiments of Tokunaga et al.\cite{Tokunaga-2009} where two regimes of non-linear ME response were also observed with a polarization drop around the critical field of 1T and a disappearance of the polarization beyond a critical field of about 4T. 

\section{conclusion}
In this paper we have analyzed the magnetoelectric response of the rare-earth orthoferrite perovskite oxides through theoretical methods based on DFT calculations, Heisenberg and analytical models taking into account the exchange striction that induces the polarization. 
With this analysis, we have shown that the non-linear character of the magnetoelectric response of GdFeO$_3$ is coming from the fact that the G-type ordering that breaks the inversion center changes non-linearly with respect to an external applied magnetic field. 
When the applied magnetic field is along the $z$ direction, the polarization reduces down to zero at a field of 4 T, i.e., when the Gd spins are all aligned with the magnetic field in a ferromagnetic state where the exchange striction is absent.
When the applied magnetic field is along the $x$ direction, the field is parallel to the main Gd G-type spin direction such that we observe two regimes: (i) from 0 to 1 T, the polarization is not affected by the field and (ii) at 1 T the Gd spin directions change from $x$ to $z$ direction (but keeping its G-type AFM ordering), which induces a strong reduction of the polarization. From 1 to 4 T the polarization is reduced non-linearly down to 0 when the Gd becomes ferromagneticaly aligned with the field along $x$.
These two regimes and the non-linear evolution of the polarization observed for the two directions of the applied magnetic field in GdFeO$_3$ is in good agreement with the experiments such that we are confident about the validity of the developed model.

Our analysis can be generalized to other rare earth perovskites in which the polarization arises from the AFM ordering and the exchange striction effect. 
Our conclusions are also general for all materials where the AFM order breaks the inversion symmetry in the presence of two different magnetic sublattices. 
For example, Wang \textit{et. al}~\cite{Wang2015} have reported the ME response of Fe$_2$Mo$_3$O$_8$ where the AFM order breaks the inversion symmetry through the exchange striction effect. The ME behavior is similar to what we have for GdFeO$_3$, i.e. giant and non-linear with similar shapes of the polarization versus magnetic field curves.
Additionally, for the systems in which it is the FM order that breaks the inversion symmetry, the same analysis can be done but instead of having a non-linear response, we will have a linear response that will be observed.

\section*{Acknowledgements}
This work has been funded by the Communaut\'e Fran\c{c}aise de Belgique (ARC AIMED G.A. 15/19-09).
EB and AS thanks the FRS-FNRS for support. J.\'I. thanks the support of the Luxembourg National Research Fund through Grant No. FNR/C18/MS/12705883/REFOX.
The authors acknowledge the CECI supercomputer facilities funded by the F.R.S-FNRS (Grant No. 2.5020.1), the Tier-1 supercomputer of the F\'ed\'eration Wallonie-Bruxelles funded by the Walloon Region (Grant No. 1117545) and the OFFSPRING PRACE project.

\bibliography{ME_Res}

\end{document}